\documentclass[10pt,a4paper]{article}
\usepackage{amsmath}
\usepackage{graphicx}
\usepackage{subcaption}
\usepackage{amssymb}
\usepackage{epstopdf}
\usepackage{cite}
\usepackage{cancel}
\usepackage{authblk}
\usepackage{hyperref}
\usepackage[english]{babel}
\hypersetup{
	colorlinks=true,
	linkcolor={blue},
	citecolor={blue},
	urlcolor={blue}
}

\DeclareMathOperator{\dd}{{\rm d}\!}
\DeclareMathOperator{\half}{\frac{1}{2}}
\newcommand{\rec}[1]{\frac{1}{#1}}
\newcommand{\pdf}[2]{\frac{\partial #1}{\partial #2}}

\title{Short Time Angular Impulse Response of Rayleigh Beams}
\author[1]{Bidhayak Goswami}
\author[2]{K. R. Jayaprakash}
\author[1]{Anindya Chatterjee}
\affil[1]{Mechanical Engineering, IIT Kanpur}
\affil[2]{Mechanical Engineering, IIT Gandhinagar}
\begin{document}
	\maketitle
	
\begin{abstract}
In the dynamics of linear structures, the impulse response function is of fundamental interest. In some cases one examines the short term response wherein the disturbance is still local and the boundaries have not yet come into play, and for such short-time analysis the geometrical extent of the structure may be taken as unbounded. Here we examine the response of slender beams to angular impulses. The Euler-Bernoulli model, which does not include rotary inertia of cross sections, predicts an unphysical and unbounded initial rotation at the point of application. A finite length Euler-Bernoulli beam, when modeled using finite elements, predicts a mesh-dependent response that
shows fast large-amplitude oscillations setting in very quickly. The simplest introduction of rotary inertia yields the Rayleigh beam model, which has more reasonable behavior including a finite wave speed at all frequencies. If a Rayleigh beam is given an impulsive moment at a location away from its boundaries, then the predicted behavior has an instantaneous finite jump in local slope or rotation, followed by smooth evolution of the slope for a finite time interval until reflections arrive from the boundary, causing subsequent slope discontinuities in time. We present a detailed study of the angular impulse response of a simply supported Rayleigh beam, starting with dimensional analysis, followed by modal expansion including all natural frequencies, culminating with an asymptotic formula for the short-time response. The asymptotic formula is obtained by breaking the series solution into two parts to be treated independently term by term, and leads to a polynomial in time. The polynomial matches the response from refined finite element (FE) simulations.
\end{abstract}

\section{Introduction}

Beams are ubiquitous structural members in engineering, especially civil and mechanical engineering, and also encountered in applied physics, micro- and nano-mechanics. The mathematical study of these flexural members dates back to the 16th century \cite{timoshenko2003strength} in the works of Jacob Bernoulli, followed by Euler. The basic assumptions with regard to the deformation kinematics, warp-free planes, planes perpendicular to the neutral axis, etc., come from the works of Euler and Bernoulli. The Euler-Bernoulli beam model is taught and studied widely owing to its simplicity and applicability for beams where either the length or the wavelength (in case of traveling waves) is large compared to the radius of gyration of the beam cross section. Many engineering beam structures obey this theory for many types of loading. One drawback of the Euler-Bernoulli beam is that the phase speed $c_{p}\propto\sqrt{\omega}$ is unbounded for large $\omega$, where $\omega$ is the frequency of the wave \cite{graff2012wave}. A slightly more advanced beam model was proposed by Lord Rayleigh \cite{rayleigh1945tos}, who added rotary inertia of cross sections to the Euler-Bernoulli beam model without adding new field variables. Unlike the Timoshenko beam which adds shear deformations \cite{timoshenko1921model,timoshenko1922model}, the Rayleigh beam model is almost as easily tractable as the Euler-Bernoulli beam and has only one field variable, namely the transverse displacement $u(x,t)$. For the Rayleigh beam model, $c_{p}$ approaches a finite limit as $\omega\rightarrow \infty$ \cite{graff2012wave}.

In this paper we take up the Rayleigh beam model because it is the simplest classical beam model that predicts {\em bounded} responses to an instantaneous {\em angular} impulse, i.e., an impulsive moment concentrated in both space and time. As we will show below, the response has several aspects of fundamental academic interest.
Our results have practical value as well, because the impulse response function (IRF) forms the basis for evaluating the response to any arbitrary excitation using a convolution integral \cite{meirovitch1997vibrations}.

A basic element of the analytical treatment for such problems lies in assuming that boundaries are far enough away so that reflections of sufficient strength do not arrive too quickly, and an analysis of the short term response can be carried out by assuming either an unbounded structure or a finite structure with analytically convenient boundary conditions. A study of the response of an Euler-Bernoulli beam to a {\em linear} impulse gives interesting bounded results \cite{chatterjee2004short}. However, the Euler-Bernoulli beam model predicts unbounded responses to {\em angular} impulses. This is the motivation for taking up the Rayleigh beam for study in this paper: it is the simplest model that incorporates rotary inertia and gives physically plausible results.

While our study is theoretical and limited to an ideal concentrated impulsive moment on an ideal Rayleigh beam, it will shed useful light on some practical situations as well. For example, the response of the beam to a localized angular impulse may be useful in computing the short time structural behavior upon the rapid stopping of a motor mounted on a beam with long span.

As indicated above, the transient response of beams excited by {\em linear} impulses (as opposed to angular impulses) is a well researched topic. A representative review of the literature follows.
An early and clever article by Zener \cite{zener1941intrinsic} studied the response of a thin plate subjected to a linear impulse for durations short enough that the effect of waves reflected from the boundaries can be neglected. Schwieger \cite{schwieger1965simple,schwieger1970central} adapted Zener's analysis to an Euler-Bernoulli beam, found the now well known $\sqrt{t}$ behavior in the linear impulse response, and conducted an experimental investigation that bore out the approximation. The detailed responses of an infinite beam as well as a simply supported beam were studied by Chatterjee \cite{chatterjee2004short} (2004), and some time later a large part of that work was developed independently by Meijaard \cite{meijaard2007lateral} (2007). The $\sqrt{t}$ nature of the response of the infinite beam was used in computing detailed ball-impact responses in \cite{bhattacharjee2018transverse}. A symbolic calculation for plates subjected to impulses was presented by Claeyssen et al.\ \cite{claeyssen2002impulse} (2002). Numerous finite element (FE) studies of beam responses are available. Roy et al.\ \cite{roy1995transient} (1995) used an FE model of a beam and investigated the short time response under viscous damping with the impact at different locations along the beam. Impulse response studies have been carried out for more complicated structural members, accounting for viscoelastic layers \cite{barkanov2000transient} and composites \cite{jayaprakash2013fatigue}. There are also papers on related topics like vibro-impacting beams \cite{wagg1999experimental} and vibration-dominated impacts \cite{bhattacharjee2020restitution}. Finally, digressing from transversely acting linear impulse loading, Kenny et al.\ \cite{kenny2000dynamic} (2000) studied the dynamic buckling of slender beams subjected to axial impulse loading and validated the results using FE analysis. 

As the foregoing literature review indicates, there are many studies that examine the behavior of beams subjected to {\em linear} impulses, but not {\em angular} impulses. In this paper we will study the transient behaviour of a beam subjected to an impulsive moment. We will have to abandon the Euler-Bernoulli beam model because its lack of rotary inertia leads to unbounded rotations; and we will take up instead the Rayleigh beam model, which has the same kinematics and strain energy, but in which the kinetic energy includes a contribution from the rotations of cross sections.

\section{Euler-Bernoulli beams under impulsive moments}

\begin{figure}[h]
	\centering
	\includegraphics[scale=0.6]{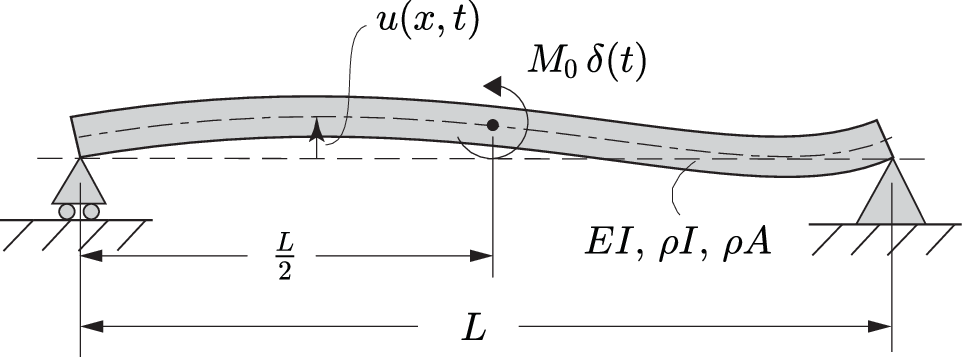}
	\caption{A slender beam subjected to a impulsive moment applied at the midpoint.}
	\label{beam_diagram}
\end{figure} 
With reference to  Fig.\ \ref{beam_diagram}, consider a simply supported Euler-Bernoulli beam of length \(L\), flexural rigidity \(EI\), material density \(\rho\), and cross sectional area \(A\). Let this beam be subjected to an impulsive moment \(M_0\) at its midpoint. The governing equation for the transverse displacement \(u\) is
\begin{equation}\label{EB_gov_eq}
	\rho A\,u_{,tt}+EI\,u_{,xxxx}=- M_0\,\delta_{,x}\left(x-\frac{L}{2}\right)\delta(t),
\end{equation}
where $x$ is the spatial coordinate along the beam, $t$ is time, subscripts denote partial derivatives, and
$\delta(\cdot)$ denotes the Dirac delta function.

\begin{figure}[h!]
	\centering
	\includegraphics[width=\linewidth]{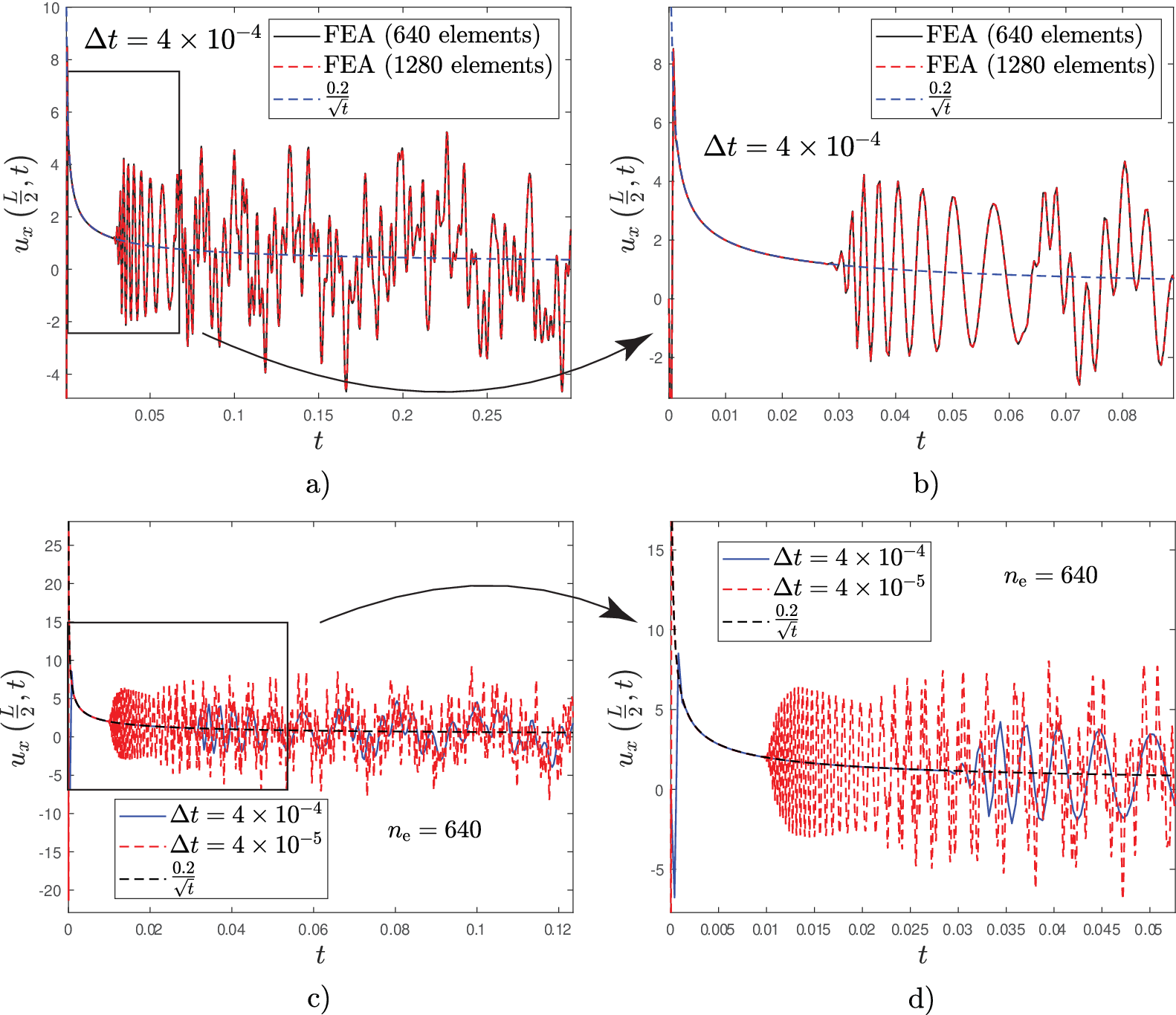}
	\caption{Computed response of a simply supported Euler-Bernoulli beam with $ L=1 $ to a central angular impulse for different mesh refinements and time steps. The 0.2 coefficient in
the $1/\sqrt{t}$ response is numerically fitted. Between subplots (a) and (b), only the time scale is changed for visibility; so also for subplots (c) and (d). Subplots (a) and (b) show that, for a fixed time step, higher mesh refinement has no effect. Subplots (c) and (d) show that, for high enough mesh refinement, the duration of the computationally obtained $1/\sqrt{t}$ behavior shrinks
as the time step is reduced. This $1/\sqrt{t}$ behavior is an artifact of implicit integration, which offers stability but damps out super-high modes, in turn removing super-quick reflections from boundaries.}
	\label{eu_unb1}
\end{figure}

Let us approach this problem using dimensional analysis \cite{langhaar1951}. The angular {\em rotation} at the impulse location, namely
$u_{,x}\left(\frac{L}{2},t\right)$, is a function of five quantities, namely \(EI\),  \(\rho A\),  \(M_0\), $L$, and \(t\). If we are interested in the short time response of the beam, we may tentatively assume that the beam length does not affect the solution because reflections of sufficient magnitude have not yet traveled back from the ends of the beam.
By this assumption, the rotation must be a dimensionless function of \(EI\),  \(\rho A\),  \(M_0\), and \(t\). These four quantities (three parameters and one variable) allow formation of a dimensionless variable, which we take to be
\begin{equation}\label{nondi_1}
	\pi_1 = (EI)^a\, (\rho A)^b\, t^c\,M_0^d.
\end{equation}
However, since the system is linear, and starts from zero initial conditions, the response must be proportional to \(M_0\), which means \(d=1\). Using routine calculations, the remaining constants must be
\begin{equation}\label{constants}
	a=-{\frac{3}{4}},\,b=-{\frac{1}{4}},\mbox{ and }\,c=-{\frac{1}{2}}.
\end{equation}
In the above, $c=-1/2$ implies that the short time response is proportional to \(1/\sqrt{t}\), which is unbounded as \(t\rightarrow 0\).
Mathematically, this unbounded response is due to non-inclusion of rotary inertia in the Euler-Bernoulli beam model. Furthermore, for a finite-length beam, the assumption that reflections from the boundary are negligible is seen to be invalid. See Fig.\ \ref{eu_unb1}. In FE simulations of a beam with unit length with large numbers of elements (we used 640 and 1280) and an implicit integration algorithm which damps out super-high frequencies, the computational result shows a short period of $1/\sqrt{t}$ behavior before high frequency oscillations appear. However, with the number of elements held fixed (at 640, which is high enough), if the time step is reduced, the duration of the $1/\sqrt{t}$ behavior shrinks as well. In the exact solution for a finite-length Euler-Bernoulli beam, reflections from the boundaries play a significant role immediately after application of the angular impulse.

With the above motivation we turn to the Rayleigh beam model, which incorporates rotary inertia but retains the simple deformation kinematics of the Euler-Bernoulli beam model \cite{rayleigh1945tos,hagedorn2007wave}. From an analytical perspective, the Rayleigh beam model is the preferred first step as compared to the Timoshenko beam model, which both includes rotary inertia and allows shear deformation. We hope that an analysis of the Timoshenko beam model may be undertaken in future work. In this paper, we focus on developing asymptotic approximations for the short time angular impulse response of Rayleigh beams.

\section{Rayleigh beams}
We now consider the same elastic beam as in Fig.\ \ref{beam_diagram}, but include rotary inertia equal to \(\rho I\) per unit length in the mathematical model. In this formulation, for an angular impulse \(M_0\) applied at the midpoint of the beam, the beam deflection \({u}(x,t)\) is governed by the equation
\begin{equation}\label{gov_eq}
	\rho A\,u_{,tt}+EI\,u_{,xxxx}-\rho I \,{u}_{,ttxx}=- M_0\,\delta_{,x}\left(x-\frac{L}{2}\right)\delta(t),
\end{equation}
where subscripts denote partial derivatives. It is interesting to note that an Euler-Bernoulli beam incorporating nonlocal effects based on Eringen's stress gradient theory \cite{eringen2002nonlocal,gopalakrishnan2017wave} assumes a similar governing equation as that of the Rayleigh beam model. However, the rotary inertia \(\rho I\) per unit length of the Rayleigh beam model is replaced by the nonlocal scaling parameter \(\rho A (e_{0}a_{\rm c})^{2}\) in a nonlocal beam, \(e_{0}\) is the nonlocal material parameter and \(a_{\rm c}\) is the internal characteristic length.

\subsection{Dimensional analysis}
Proceeding along similar lines as in the previous section, and again dropping $L$ because we are interested in the {\em short time} response, we recognize that the angular response is a function of \(M_0,\,EI,\,\rho A,\,\rho I\), and \(t\). There are now two dimensionless quantities, and we include $M_0$ in one of them and $t$ in the other, to write
\begin{equation}
	\eta_1= \left(EI\right)^{a_1}\,(\rho I)^{b_1}\,(\rho A)^{c_1}\,(M_{0})^{d_1}\mbox{ and }\eta_2= \left(EI\right)^{a_2}\,(\rho I)^{b_2}\,(\rho A)^{c_2}\,(t)^{d_2}.
\end{equation}
Setting \(d_1=d_2=1\) for definiteness, using routine methods, we obtain
\begin{equation}
	a_1=b_1=-\frac{1}{2},\,c_1=0, \mbox{ and } a_2=c_2=\frac{1}{2},b_2=-1,
\end{equation}
or
\begin{equation}
	\eta_1=\frac{M_0}{I\sqrt{E\rho}},\,\,\eta_2=t\sqrt{\frac{EA}{\rho I}}.
\end{equation}
Accordingly, the {\em short-time} rotational response at the midpoint of the beam, being dimensionless, is of the form
\begin{equation}
	u_{,x}\left(\frac{L}{2},t\right)=f\left( \frac{M_0}{I\sqrt{E\rho}}, t\sqrt{\frac{EA}{\rho I}}\right).
\end{equation}
Linearity of Eq.\ \ref{gov_eq} implies the response is proportional to \(M_0\), so we must have
\begin{equation}
\label{dim2}
	u_{,x}\left(\frac{L}{2},t\right)=\frac{M_0}{I\sqrt{E\rho}}\, f_0\left(t\sqrt{\frac{EA}{\rho I}} \right),
\end{equation}
where $f_0$ is to be determined as a function of nondimensional time \(\tau=t\sqrt{EA/\rho I}\). Since we have neglected reflections from the ends by dropping $L$ from this dimensional analysis, Eq.\ \ref{dim2} is expected to hold  for $0 < \tau \ll 1$, and $ L $ large.

Next, we consider the exact solution using a modal expansion.

\subsection{Modal solution}

It may be verified that the differential operator of the Rayleigh beam is self-adjoint \cite{hagedorn2007wave,meirovitch1997vibrations} and the eigenfunctions of the resulting eigenvalue problem are orthonormal. By direct substitution, for a simply supported beam, the eigenfunctions are seen to be pure sines. Again assuming simple supports at both  ends of the beam, and expanding the solution using the beam's eigenfunctions \cite{hagedorn2007wave}, we write
\begin{equation}\label{sol_basis}
	{u}(x,\tau)=\sum_{n=1}^{\infty} q_n(\tau) \sin \left ( \frac{n\pi x}{L} \right ),  
\end{equation}
where the time-varying coefficients \(q_n(\tau)\) satisfy \(q_n(0)=0\). 

With suitable choice of units of mass, length, and time, we can make \(\rho A=1\), \(\rho I=1\), and \(EI=1\). Now $ L $ can be treated as dimensionless. 
Substituting Eq.\ \ref{sol_basis} in Eq.\ \ref{gov_eq} yields
\begin{equation}
	\sum_{n=1}^{\infty}\sin(p_n x)\left(q''_n(1+p_n^2)+p_n^4\,q_n\right)=-\delta_{,x}\left(x-\frac{L}{2}\right)\delta(\tau)
\end{equation}
where \(p_n=n\,\pi/L\), and the prime
\((\cdot)'\) indicates a \(\tau\)-derivative.
Multiplying both sides with \(\sin(p_k x)\) for positive integers $k$, and integrating over the length of the beam as usual, we obtain
\begin{equation}\label{qk_diff}
	q''_k+\omega^2_k \,q_k=\frac{2\,p_k\,\cos\left(\frac{p_k\,L}{2}\right)}{L\,\left(1+p_k^2\right)}\delta(\tau),
\end{equation}
where  \(\omega_k={p_k^2}/\sqrt{1+p_k^2}\). For zero initial conditions, the solution of Eq.\ \ref{qk_diff} is
\begin{equation}\label{qk_sol}
	q_k(\tau)=\frac{2\,p_k\,\cos\left(\frac{k\,\pi}{2}\right)}{L\,(1+p_k^2)\,\omega_k}\sin(\omega_k\, \tau),
\end{equation}
yielding
\begin{equation}
	u(x,\tau)=\frac{2}{L}\sum_{k=1}^{\infty}\frac{\cos\left(\frac{k\,\pi}{2}\right)}{p_k\,\sqrt{1+p_k^2}}\sin(p_k x)\sin\left(\frac{p_k^2}{\sqrt{1+p_k^2}}\, \tau\right).
\end{equation}
Terms containing odd $k$ drop out of the above sum, and the rotation at \(x=L/2\) is then
\begin{align}\label{slope_fun}
	u_{,x}\left(\frac{L}{2},\tau\right)
	=\frac{2}{L}\sum_{k=1}^{\infty}\frac{1}{\sqrt{1+4\,k^2\,a^2}}\,\sin\left(\frac{4\,k^2\,a^2}{\sqrt{1+4\,k^2\,a^2}}\,\tau\right)
\end{align}
where \(a=\pi/L\).
Taking a $\tau$-derivative term by term, the angular velocity at \(x=L/2\) is formally
\begin{equation}\label{angular_velocity_series}
	u_{,x\tau}\left(\frac{L}{2},\tau\right)
	=\frac{2}{L}\sum_{k=1}^{\infty}\frac{4\,k^2\,a^2}{1+4\,k^2\,a^2}\cos\left(\frac{4\,k^2\,a^2}{\sqrt{1+4\,k^2\,a^2}}\,\tau\right),
\end{equation}  
but the series in Eq.\ \ref{angular_velocity_series} diverges at $\tau=0$. We need more careful analysis.

\begin{figure}[ht]
	\centering
	\includegraphics[scale=0.8]{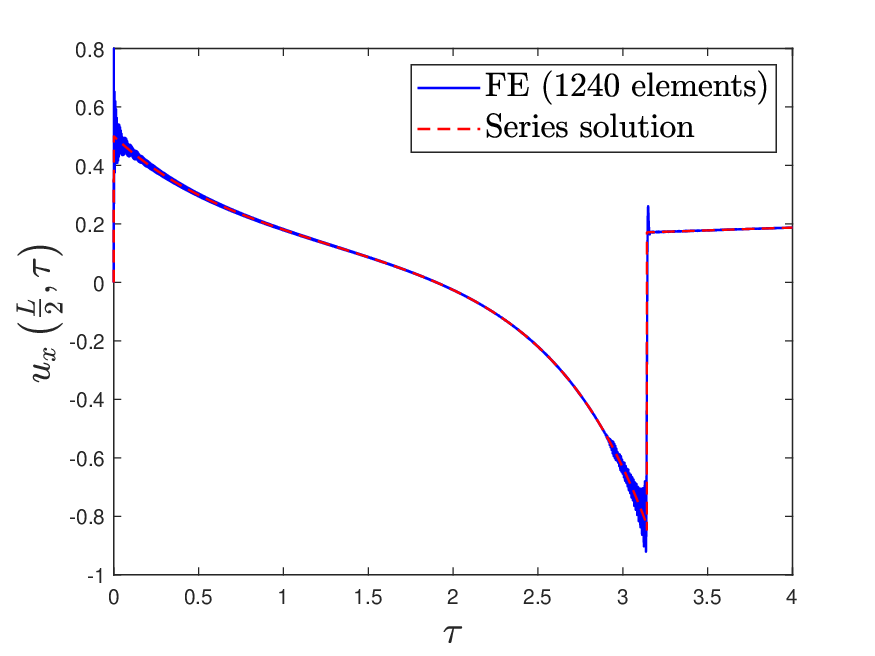}
	\caption{Angular impulse response at the midpoint of a Rayleigh beam of length \(L=\pi\) calculated with a step size \(h=4\times10^{-4}\), from an FE model with 1280 elements and the series given in Eq.\ \ref{slope_fun} summed to \(10^5\) terms. The series solution shows discontinuities at $\tau=0$ and $\tau = \pi$. The discontinuity at \(\tau=\pi\) is because of the  arrival of high frequency reflections from the ends of the beam. The high frequency oscillations in the FE solution are numerical artifacts (see the main text).}
	\label{ang_imp_res}
\end{figure}

The series solution in Eq.\ \ref{slope_fun} for the slope at the center of the beam of length \(\pi\) exhibits a discontinuity at \(\tau=0\) (shown in Fig.\ \ref{ang_imp_res}), but the behavior is bounded unlike that of the Euler-Bernoulli beam. The discontinuity arrives some time later at points slightly separated from the point of application of the moment impulse.
Further, the response is discontinuous at \(\tau=\pi\) as well. This is because of two reasons. First, for a beam length $\pi$, the travel distance from the center of the beam to either end and back is $\pi$. Second, with physical parameters set to unity as above, the phase velocity as a function of frequency $\omega$ approaches unity from below as $\omega \rightarrow \infty$. For this reason, infinitely many frequencies return after reflection as $\tau \rightarrow \pi$. We evaluate the series solution by summing \(10^5\) terms, and demonstrate the first two discontinuities.
The series for longer times shows subsequent discontinuities at integer multiples of $\pi$ (not displayed for brevity).

Finite element (FE) analysis of a Rayleigh beam with 1280 fixed-length elements (see details in {\bf Appendix} \ref{FEA}) yields results that match the series solution
including the discontinuity at $\tau = \pi$, except for some high-frequency oscillations near the discontinuities. We believe the oscillations are artifacts of the level of mesh refinement, the chosen time step, and the implicit integration algorithm used. The analytical series solution is correct in principle, but should be verified by an independent calculation: the FE solution, which matches over a large region, provides that verification.

Equation \ref{gov_eq} is a linear PDE which has been solved using a series expansion. In Fig. \ref{ang_imp_res}, the response of a beam of length \(\pi\) has been plotted to show the arrival of reflected waves from the boundary. However,  in order to get physically more meaningful results, we should consider \(L\gg 1\). This is because both \(\rho A\) and \(\rho I \) have been set to unity above, and so the lateral dimensions of the beam are of \({\cal O}(1)\). Hence, we must choose \(L\) large compared to unity so that the assumptions of a slender beam are valid.

\begin{figure}[h!]
	\centering
	\includegraphics[scale=0.58]{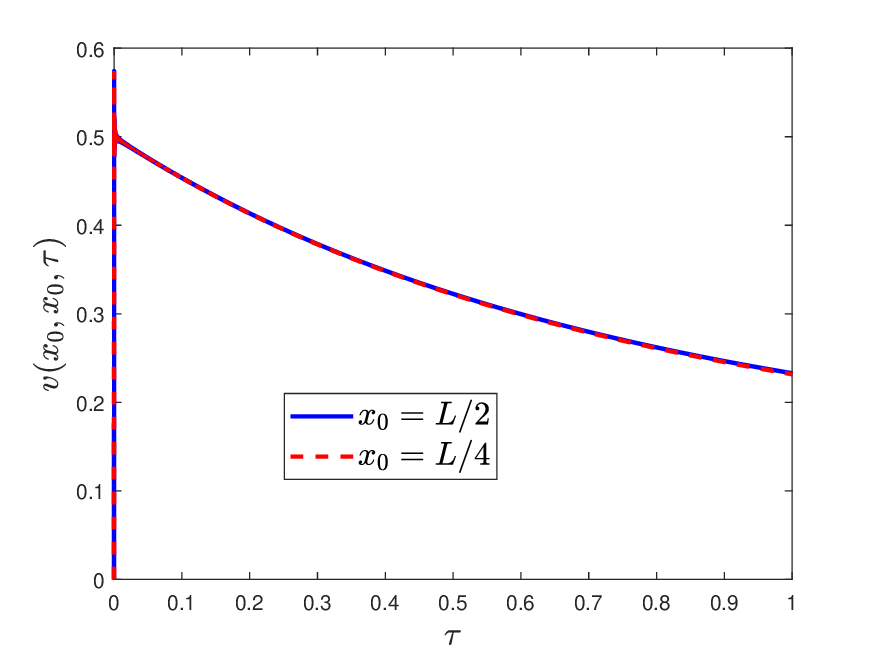}
	\caption{Responses \(v(x_0,x_0,\tau)\) of a Rayleigh beam with  \(L=12\) for \(x_0=L/2\) and \(x_0=L/4\). The two responses are clearly the same.}
	\label{ST_diff_pts}
\end{figure}

For beams with  \(L\gg 1\), the short time impulse response is independent of the point of application of the load.
Let us denote the rotational response at a point \(x\) at time \(\tau\) due to an angular impulse applied at point \(x_0\) by \(v(x,x_0,\tau)\). Figure \ref{ST_diff_pts} shows the responses of a beam of length \(L=12\) subjected to an angular impulse at  \(x=L/2\) and at \(x=L/4\).

Figure \ref{ST_diff_pts} establishes that the short time response is independent of the point of application of the load provided the beam is somewhat long and the load is not applied too close to one of the ends. 
For subsequent calculations, we choose the midpoint of the beam for  analytical convenience because then half  the terms drop out of the  series solution (Eq.\ \ref{slope_fun}). The series solution of Eq.\ \ref{slope_fun} must now be treated analytically to obtain an asymptotic description of its behaviour for small \(\tau\).

We now come to the main academic research contribution of this paper. We will show below that
\begin{multline}
\sum_{k=1}^{\infty}\frac{1}{\sqrt{1+4\,k^2\, a^2}}\sin\left(\frac{4\,k^2\,a^2}{\sqrt{1+4\,k^2\,a^2}}\tau\right)=
{\frac {\pi}{4\,a}}-{\frac {\pi\,{\rm coth} \left({\frac {\pi}{2\,a}}\right)
		}{4\,a}}\,\tau+{\frac {3\pi}{16\,a}}\,{\tau}^{2}+{\frac {\pi \left(\pi -5\,a\sinh
		\left( \frac{\pi}{a} \right) \right) }{48\,a^2\left(\cosh \left( \frac{\pi}{a}
		\right)-1\right) }}\,{\tau}^{3}+\mathcal{O}\left(\tau^{4}\right),\\
	    a=\frac{\pi}{L},\mbox{ and }0<\tau\ll 1.
\end{multline}
\section{Short time asymptotic solution}
We begin with the infinite sum in Eq.\ \ref{slope_fun},
\begin{equation}\label{res}
	S(\tau)=\,\sum_{k=1}^{\infty}\frac{1}{\sqrt{1+4\,k^2\,a^2}}\,\sin\left(\frac{4\,k^2\,a^2}{\sqrt{1+4\,k^2\,a^2}}\,\tau\right)
	\mbox{ for }0<\tau\ll 1.
\end{equation}
It is an anharmonic series, and if differentiated term by term it diverges.
Let us consider a large number $N\approx 1/\sqrt{\tau}$, and split $S(\tau)$ as follows 
\begin{equation}\label{2sum}
	S(\tau)={\sum_{k=1}^{N}f(k,\tau)}+{\sum_{k=N+1}^{\infty}f(k,\tau)},\quad{\rm where }\,f(k,\tau)=\frac{1}{\sqrt{1+4\,k^2\,a^2}}\sin\left(\frac{4\,k^2\,a^2}{\sqrt{1+4\,k^2\,a^2}}\,\tau\right).
\end{equation}
We now state a basic result which we will use in our approximations. A formal proof is given in {\bf Appendix}~\ref{app1}.
\smallskip

\noindent{\bf Lemma 1}
	{\em Consider a function $g(y):\mathbb{R}\rightarrow\mathbb{R}$. Let all its derivatives tend to $0$ as $y\rightarrow \infty$, with higher derivatives being asymptotically smaller than lower derivatives as \(y\rightarrow\infty\); and assume further that the sum $\sum_{k=N+1}^{\infty}g(k)$ and the integral
$\int_{N}^{\infty}g(y)\;{\rm d}y$
exist. Then}
\begin{equation} \label{Lem1}
 \sum_{k=N+1}^{\infty}g(k) = \int_{N}^{\infty}g(y)\;{\rm d}y-\frac{g(N)}{2}-\frac{g'(N)}{12}+\frac{g'''(N)}{720} - \frac{g^{(5)}(N)}{30240} +
\frac{g^{(7)}(N)}{1209600} \cdots \end{equation}
%
This result will be used repeatedly below. The known sums of some infinite series are given in {\bf Appendix} \ref{app2}, which will be used below as well.

We now treat the two terms in the RHS of Eq.\ \ref{2sum} separately.
\subsection{First term of Eq.\ \ref{2sum}}
Considering the first sum on the right hand side of Eq.\ \ref{2sum}, the argument of the `sin' function, namely $\frac{4\,k^2\,a^2}{\sqrt{1+4\,k^2\,a^2}}\,\tau$, is small for all $k < N$ for $\tau \ll 1$. Thus, termwise Taylor series expansion is allowed, and it yields 
\begin{equation}\label{t1_taylor}
	S_1(\tau)=\sum_{k=1}^{N}f(k,\tau)={\sum_{k=1}^{N}{\frac {4\,{k}^{2}\,a^2\,\tau}{4\,{k}^{2}\,a^2+1}}}-{\sum_{k=1}^{N}{\frac {32\,{k}^{6}\,a^6\,{\tau}^{3}}{3\,
				\left( 4\,{k}^{2}\,a^2+1 \right) ^{2}}}}+{\sum_{k=1}^{N}{\frac {128\,{k}^{10}\,a^{10 }\,{\tau}^{5}}{15\,
				\left( 4\,{k}^{2}\,a^2+1 \right) ^{3}}}}+{\rm h.o.t}.
\end{equation}
The first term on the right hand side of Eq.\ \ref{t1_taylor} can be written as
\begin{align}\label{T1a}
	{ S_{1,1}}=
	\sum_{k=1}^{N}\frac {4\,{k}^{2}\,a^2\,\tau}{4\,{k}^{2}\,a^2+1}&=N\tau-\tau\sum_{k=1}^{N}\frac{1}{4\,k^2\,a^2+1}=N\tau-\tau\sum_{k=1}^{\infty}\frac{1}{4\,k^2\,a^2+1}+\tau\sum_{k=N+1}^{\infty}\frac{1}{4\,k^2\,a^2+1},
\end{align}
and therefore  (see {\bf Appendix} \ref{app2})
\begin{equation}
\label{T1} {S_{1,1}}= N\tau+ \tau\sum_{k=N+1}^{\infty}\frac{1}{4\,k^2\,a^2+1}
	-\left({\frac {\pi\,{\rm coth} \left({\frac {\pi}{2\,a}}\right)}{4\,a}}-{\frac{1}{2}}\right)\tau .
\end{equation}
By Lemma 1, we can write
\begin{align}
	\sum_{k=N+1}^{\infty}\frac{1}{1+4\,k^2\,a^2}& \sim \int_{N}^{\infty}\frac{\dd x}{1+4\,x^2\,a^2}-{\frac {1}{8\,{N}^{2}{a}^{2}+2}}+{\frac {2\,N{a}^{2}}{3\, \left( 4\,{
				N}^{2}{a}^{2}+1 \right) ^{2}}}-{\frac {64\,{N}^{3}{a}^{6}}{15\,
			\left( 4\,{N}^{2}{a}^{2}+1 \right) ^{4}}}+{\frac {8\,N{a}^{4}}{15\,
			\left( 4\,{N}^{2}{a}^{2}+1 \right) ^{3}}}
	\nonumber\\
	&=\frac{\pi-2\,\tan^{-1}(2Na)}{4\,a}
	-{\frac {1}{8\,{N}^{2}{a}^{2}+2}}+{\frac {2\,N{a}^{2}}{3\, \left( 4\,{
				N}^{2}{a}^{2}+1 \right) ^{2}}}-{\frac {64\,{N}^{3}{a}^{6}}{15\,
			\left( 4\,{N}^{2}{a}^{2}+1 \right) ^{4}}}+{\frac {8\,N{a}^{4}}{15\,
			\left( 4\,{N}^{2}{a}^{2}+1 \right) ^{3}}}
\end{align}
Substituting $N=1/\sqrt{\tau}$ and expanding in a series for small \(\tau\),
\begin{multline}\label{T1_int}
	\sum_{k=N+1}^{\infty}\frac{1}{1+4\,k^2\,a^2} \approx
	{\frac {1}{4\,{a}^{2}}\sqrt {\tau}}-{\frac {\tau}{8\,{a}^{2}}}+ \left( -{
		\frac {1}{48\,{a}^{4}}}+{\frac {1}{24\,{a}^{2}}} \right) {\tau}^{{\frac{3
			}{2}}}+{\frac {{\tau}^{2}}{32\,{a}^{4}}}+ \left( {\frac {1}{320\,{a}^{6}}
	}-{\frac {1}{48\,{a}^{4}}}-{\frac {1}{120\,{a}^{2}}} \right) {\tau}^{{
			\frac{5}{2}}}\\-{\frac {{\tau}^{3}}{128\,{a}^{6}}}+O \left( {\tau}^{{\frac{7}{
				2}}} \right)
\end{multline}
Substituting Eq.\ \ref{T1_int} in Eq.\ \ref{T1}, we obtain
\begin{equation}
	{S_{1,1}}=
	\sum_{k=1}^{N}\frac {4\,{k}^{2}\,a^2\,\tau}{4\,{k}^{2}\,a^2+1}\approx
	\sqrt {\tau}+ \left( -{\frac {\pi\,{\rm coth} \left({\frac {\pi}{2\,a}}
			\right)}{4\,a}}+{\frac{1}{2}} \right) \tau+{\frac {{\tau}^{{\frac{3}{2}}}}{4\,{a}^{2}}}-{\frac {{\tau}^{2}}{8\,{a}^{2}
		}}+ \left( -{\frac {1}{48\,{a}^{4}}}+{\frac {1}{24\,{a}^{2}}} \right) 
	{\tau}^{{\frac{5}{2}}}+{\frac {{\tau}^{3}}{32\,{a}^{4}}}+O \left( {\tau}^{{
	\frac{7}{2}}} \right)
\end{equation}
Next, the second term on the right hand side of Eq.\ \ref{t1_taylor}, using partial fractions, takes the form
\begin{equation}
	{S_{1,2}}=
	-\sum_{k=1}^{N}{\frac {32\,{k}^{6}\,a^6\,{\tau}^{3}}{3 \left( 4\,{k}^{2}\,a^2+1 \right) ^{2}}}=\tau^3
	\sum_{k=1}^{N}\left(-{\frac {2{k}^{2}\,a^2}{3}}+{\frac{1}{3}}- \frac{1}{\left( 8\,{k}^{2}\,a^2+2 \right)}+{\frac {1}{6\left( 4\,{k}^{2}\,a^2+1 \right) ^{2}}}
	\right),
\end{equation}
which yields,
\begin{multline}
	{S_{1,2}}=\tau^3\left({\frac {2\,N}{9} \left( {\frac{3}{2}}- \left( {N}^{2}+{\frac {3\,N}{
				2}}+{\frac{1}{2}} \right) {a}^{2} \right) }
	+\sum_{k=1}^{\infty}\left(-\frac{1}{\left( 8\,{k}^{2}\,a^2+2 \right)}+{\frac {1}{6\left( 4\,{k}^{2}\,a^2+1 \right) ^{2}}}\right)\right)\\
	-\tau^3\sum_{k=N+1}^{\infty}\left(-\frac{1}{\left( 8\,{k}^{2}\,a^2+2 \right)}+{\frac {1}{6 \left( 4\,{k}^{2}\,a^2+1 \right) ^{2}}}\right),
\end{multline}
which further yields (see {\bf Appendix}~\ref{app2})

\begin{multline}\label{T2_exp}
	{S_{1,2}}=\tau^3\left({\frac {2\,N}{9} \left( {\frac{3}{2}}- \left( {N}^{2}+{\frac {3\,N}{
				2}}+{\frac{1}{2}} \right) {a}^{2} \right) }+ 
	{\frac {1}{96\,{a}^{2}} \left(  \left( {\rm coth} \left({\frac {\pi}{2
				\,a}}\right) \right) ^{2}{\pi}^{2}+16\,{a}^{2}-10\,\pi\,{\rm coth} 
		\left({\frac {\pi}{2\,a}}\right)a-{\pi}^{2} \right) }
	  \right)\\
	-\tau^3\sum_{k=N+1}^{\infty}\left(-\frac{1}{\left( 8{k}^{2}\,a^2+2 \right)}+{\frac {1}{6 \left( 4{k}^{2}\,a^2+1 \right) ^{2}}}\right).
\end{multline}
Using Lemma 1 (Eq.\ \ref{Lem1}), we can approximate the sum after evaluating the integral,

\begin{multline}\label{T2_int}
	\sum_{k=N+1}^{\infty}-\frac{1}{\left( 8{k}^{2}\,a^2+2 \right)}+{\frac {1}{6 \left( 4{k}^{2}\,a^2+1 \right) ^{2}}} =
	-{\frac {20\,{N}^{2}\pi\,{a}^{2}-40\,{N}^{2}\arctan \left( 2\,Na
			\right) {a}^{2}+4\,Na+5\,\pi-10\,\arctan \left( 2\,Na \right) }{
			\left( 192\,{N}^{2}{a}^{2}+48 \right) a}}\\
		+ \frac{1}{\left( 16\,{N}^{2}{a}^{2}+
			4 \right)}-{\frac {1}{12\, \left( 4\,{N}^{2}{a}^{2}+1 \right) ^{2
	}}}-{\frac {4\,N{a}^{2}}{3\, \left( 8\,{N}^{2}{a}^{2}+2 \right) ^{2}}}
	+{\frac {2\,N{a}^{2}}{9\, \left( 4\,{N}^{2}{a}^{2}+1 \right) ^{3}}}+\cdots	
\end{multline}
Substituting Eq.\ \ref{T2_int} in Eq.\ \ref{T2_exp} followed by letting $N=1/\sqrt{\tau}$, and expanding for small \(\tau\) yields
\begin{equation}
	{S_{1,2}} =
	-{\frac {2\,{a}^{2}}{9}{\tau}^{{\frac{3}{2}}}}-{\frac {{a}^{2}{\tau}^{
				2}}{3}}+ \left( {\frac{1}{3}}-{\frac {{a}^{2}}{9}} \right) {\tau}^{{
			\frac{5}{2}}}+{\frac {{\tau}^{3}}{96\,{a}^{2}} \left( {\pi}^{2}
		\left( {\rm coth} \left({\frac {\pi}{2\,a}}\right) \right) ^{2}-10\,
		\pi\,{\rm coth} \left({\frac {\pi}{2\,a}}\right)a-{\pi}^{2}+16\,{a}^
		{2} \right) }+{\cal O} \left( {
		\tau}^{\frac{7}{2}} \right)
\end{equation}
The last term on the right hand side of Eq.\ \ref{t1_taylor} can be expressed in partial fractions as well, to yield
\begin{align}\label{T3_1}
	{S_{1,3}}&=
	\sum_{k=1}^{N}{\frac {128\,{k}^{10}\,a^{10}{\tau}^{5}}{15\left( 4\,{k}^{2}\,a^2+1 \right) ^{3}}}\nonumber \\
	&=\tau^5\sum_{k=1}^{N}\left(
	{\frac {2\,{a}^{4}{k}^{4}}{15}}-{\frac {{a}^{2}{k}^{2}}{10}}+{\frac{1}
		{20}}-{\frac {1}{48\,{a}^{2}{k}^{2}+12}}-{\frac {1}{120\, \left( 4\,{a
			}^{2}{k}^{2}+1 \right) ^{3}}}+{\frac {1}{24\, \left( 4\,{a}^{2}{k}^{2}
			+1 \right) ^{2}}}
	\right)\nonumber\\
	&=\tau^5\left({\frac {2\,N}{75} \left( {\frac{15}{8}}+ \left( {N}^{4}+{\frac {5\,{N}
				^{3}}{2}}+{\frac {5\,{N}^{2}}{3}}-{\frac{1}{6}} \right) {a}^{4}+
		\left( -{\frac {5\,{N}^{2}}{4}}-{\frac {15\,N}{8}}-{\frac{5}{8}}
		\right) {a}^{2} \right) }
	\right)\nonumber \\
	&\quad+\tau^5{\frac {2016\,{N}^{4}\arctan \left( 2\,a\,N \right) {a}^{4}-1008\,{N}^
			{4}\pi\,{a}^{4}-272\,{N}^{3}{a}^{3}+1008\,{N}^{2}\arctan \left( 2\,a\,
			N \right) {a}^{2}}{3840\, \left( 4\,{N}^{2}{a}^{2}+1
			\right) ^{2}a}}\nonumber \\
	&\quad+\tau^5\frac{-504\,{N}^{2}\pi\,{a}^{2}-60\,a\,N+126\,\arctan
		\left( 2\,a\,N \right) -63\,\pi}{3840\, \left( 4\,{N}^{2}{a}^{2}+1
		\right) ^{2}a}-\tau^5\sum_{k=1}^\infty \frac{1}{12\left(4\,k^2\,a^2+1\right)}\nonumber\\
	&\quad -\tau^5\sum_{k=1}^\infty \frac{1}{120\left(4\,k^2\,a^2+1\right)^3}+\tau^5\sum_{k=1}^\infty \frac{1}{24\left(4\,k^2\,a^2+1\right)^2}+\tau^5\sum_{k=N+1}^\infty \frac{1}{12\left(4\,k^2\,a^2+1\right)}\nonumber\\
	&\quad \tau^5\sum_{k=N+1}^\infty \frac{1}{120\left(4\,k^2\,a^2+1\right)^3}-\tau^5\sum_{k=N+1}^\infty\frac{1}{24\left(4\,k^2\,a^2+1\right)^2}.
\end{align}
Again using Lemma 1, evaluating the relevant integrals, considering $N=1/\sqrt{\tau}$, and expanding for small \(\tau\), we obtain
\begin{equation}\label{T3}
	S_{1,3}\approx {\frac {2\,{a}^{4}}{75}{\tau}^{{\frac{5}{2}}}}+{\frac {{a}^{4}{\tau}^{
				3}}{15}}
	+\mathcal{O}\left(\tau^{\frac{7}{2}}\right).
\end{equation}
Finally, the first term of Eq.\ \ref{2sum} for small $\tau$ is
\begin{multline}\label{S1}
	S_1(\tau)\approx S_{1,1}+S_{1,2}+S_{1,3}=\sqrt {\tau}+{\frac {\tau}{4\,a} \left( -\pi\,{\rm coth} \left({\frac {\pi}{
				2\,a}}\right)+2\,a \right) }+ \left( {\frac {1}{4\,{a}^{2}}}-{\frac {2
			\,{a}^{2}}{9}} \right) {\tau}^{{\frac{3}{2}}}+ \left( -{\frac {1}{8\,{a}^
			{2}}}-{\frac {{a}^{2}}{3}} \right) {\tau}^{2}\\
		+ \left( -{\frac {1}{48\,{a}
			^{4}}}+{\frac {1}{24\,{a}^{2}}}+{\frac{1}{3}}-{\frac {{a}^{2}}{9}}+{
		\frac {2\,{a}^{4}}{75}} \right) {\tau}^{{\frac{5}{2}}}\\
	 + \left( {\frac {1
		}{32\,{a}^{4}}}+{\frac {1}{96\,{a}^{2}} \left(  \left( {\rm coth} 
		\left({\frac {\pi}{2\,a}}\right) \right) ^{2}{\pi}^{2}-10\,{\rm coth} 
		\left({\frac {\pi}{2\,a}}\right)\pi\,a-{\pi}^{2}+16\,{a}^{2}
		\right) }+{\frac {{a}^{4}}{15}} \right) {\tau}^{3}+{\cal O}\left(\tau^{\frac{7}{2}}\right)	
\end{multline}

\subsection{Second term of Eq.\ \ref{2sum}}
Considering now the second term of Eq.\ \ref{2sum}, we will again write it as an integral plus a sum of discrete terms (Lemma 1).
For the integral, writing $y$ in place of $k$, treating $y$ as large and $\tau$ as small, we obtain
\begin{multline}\label{sn2}
	\frac{1}{\sqrt{1+4\,y^2\,a^2}}\sin\left( \frac{4y^2\,a^2}{\sqrt{1+4\,y^2\,a^2}} \tau\right)\approx
	{\frac {\sin \left( 2\,a\,\tau\,y \right) }{2\,a\,y}}-{\frac {\cos
			\left( 2\,a\,\tau\,y \right) \tau}{8\,{a}^{2}{y}^{2}}}+\frac{1}{y^3}\left(-{\frac {\sin \left( 2\,a\,\tau\,y \right) }{16\,{a}^{3}}}-{\frac {
			\sin \left( 2\,a\,\tau\,y \right) {\tau}^{2}}{64\,{a}^{3}}}
		\right)\\
		+\frac{1}{y^4}\left({\frac {\cos \left( 2\,a\,\tau\,y \right) \tau}{64\,{a}^{4}}}+{\frac {
				\cos \left( 2\,a\,\tau\,y \right) \tau\, \left( {\tau}^{2}+18 \right) 
			}{768\,{a}^{4}}}
		\right)\\
		+\frac{1}{y^5}\left({\frac {3\,\sin \left( 2\,a\,\tau\,y \right) }{256\,{a}^{5}}}+{\frac {
				\sin \left( 2\,a\,\tau\,y \right) {\tau}^{2}}{512\,{a}^{5}}}+{\frac {
				\sin \left( 2\,a\,\tau\,y \right) {\tau}^{2} \left( {\tau}^{2}+72
				\right) }{12288\,{a}^{5}}}
		\right)+\mbox{h.o.t}.
\end{multline}
The first term in the RHS of Eq.\ \ref{sn2}, when integrated over \([N,\infty)\), can be expressed as
\begin{equation}
	\int_{N}^\infty {\frac {\sin \left( 2\,a\,\tau\, y \right) }{2\,a\,y}}\dd y =\int_{0}^{\infty}{\frac {\sin \left( 2\,a\,\tau \,y \right) }{2\,a\,y}}\dd y-\int_{0}^N
	{\frac {\sin \left( 2\,a\,\tau\, y \right) }{2\,a\,y}}\dd y =\frac{\pi}{4\,a}-\int_{0}^N
	{\frac {\sin \left( 2\,a\,\tau\, y \right) }{2\,a\,y}}\dd y. \label{fun}
\end{equation}
A trick used in Eq.\ \ref{fun} may be noted. Although the integrand has been obtained using a large-$y$ expansion, as far as the integral itself goes, Eq.\ \ref{fun}
is exact. In Eq.\ \ref{fun}, $y$ is simply a dummy variable of integration. In particular, the last integral uses small values of the integration variable.
In that final integral, since $y\tau$ is small for all $y\in (0,N)$, series expansion of $\sin(2ay\tau)/{2ay}$ and then term by term integration is possible (details omitted
for brevity). The rest of the terms in the right hand side of Eq.\ \ref{sn2} can be routinely integrated by parts to obtain asymptotic approximations (see page 252 of Bender and Orszag \cite{bender1999advanced} for examples)

Accordingly, the integral on the right hand side of Eq.\ \ref{sn2} with the limits $N=1/\sqrt{\tau}$ to $\infty$ results in
\begin{multline}\label{s2int}
	\int_{\rec{\sqrt{\tau}}}^{\infty}\frac{1}{\sqrt{1+4y^2\,a^2}}\sin\left( \frac{4\,a^2\,y^2}{\sqrt{1+4\,a^2\,y^2}} \tau\right)\dd y\approx
	{\frac {\pi}{4\,a}}-\sqrt {\tau}-{\frac {-1981808640\,{a}^{12}+2229534720
			\,{a}^{8}}{8918138880\,{a}^{10}}{\tau}^{{\frac{3}{2}}}}+{\frac {3\,\pi\,{
				\tau}^{2}}{16\,a}}\\
			-{\frac {1}{8918138880\,{a}^{10}} \left( 2972712960\,{a
		}^{10}-185794560\,{a}^{6}+{\frac {1189085184\,{a}^{14}}{5}} \right) {\tau
		}^{{\frac{5}{2}}}}+{\cal O}\left(\tau^{\frac{7}{2}}\right).
\end{multline}
The second term of Eq.\ \ref{2sum} requires us to subtract some discrete terms as well:
\begin{multline}
	\left(\half f(N,\tau)+\rec{12}\pdf{f(y,\tau)}{y}\bigg\rvert_{y=N}-\frac{1}{720}\frac{\partial^3 f(x,\tau)}{\partial x^3}\bigg\rvert_{x=N}\right)\bigg\rvert_{N=\rec{\sqrt{\tau}}}=
	{\frac {\tau}{2}}-{\frac {{\tau}^{2} \left( 480\,{a}^{4}+180
			\right) }{1440\,{a}^{2}}}-{\frac {160\,{a}^{4}-60}{1440\,{a}
			^{2}}{\tau}^{{\frac{5}{2}}}}\\
		-{\frac {{\tau}^{3} \left( -96\,{a}^{8}-
			240\,{a}^{4}-45 \right) }{1440\,{a}^{4}}}+{\cal O}\left(\tau^{\frac{7}{2}}\right).
\end{multline}
Hence,
\begin{multline}
	S_2(\tau)\approx
	{\frac {\pi}{4\,a}}-\sqrt {\tau}-{\frac {\tau}{2}}+{\frac {3}{16\,{a}^{4}}
		\left( {\frac {32\,{a}^{6}}{27}}-{\frac {4\,{a}^{2}}{3}} \right) {\tau}^
		{{\frac{3}{2}}}}+{\frac {3\,{\tau}^{2}}{16\,{a}^{4}} \left( {\frac {16\,{
					a}^{6}}{9}}+\pi\,{a}^{3}+{\frac {2\,{a}^{2}}{3}} \right) }\\
				+{\frac {-
			18816\,{a}^{10}+78400\,{a}^{8}-235200\,{a}^{6}-29400\,{a}^{4}+14700\,{
				a}^{2}}{705600\,{a}^{6}}{\tau}^{{\frac{5}{2}}}}+{\frac {3\,{\tau}^{3}}{16\,{
				a}^{4}} \left( -{\frac {16\,{a}^{8}}{45}}-{\frac {8\,{a}^{4}}{9}}-{
			\frac{1}{6}} \right) }+{\cal O}\left(\tau^{\frac{7}{2}}\right).
\end{multline}
We now have both terms needed for Eq.\ \ref{2sum}. Adding them up,
Eq.\ \ref{res} for small $\tau$ yields
\begin{equation}
	S(\tau)={\frac {\pi}{4\,a}}-{\frac {\pi}{4\,a}{\rm coth} \left({\frac {\pi
			}{2\,a}}\right)\tau}+{\frac {3\,\pi\,{\tau}^{2}}{16\,a}}+{ \frac{\left({\pi}^{2} -5\,\pi\, a\sinh \left({\frac {\pi}{a}} \right) \right)
		}{ {\cosh \left( {\frac {\pi}{a}} \right)-1 } } \frac {{\tau}^{3}}{48
		\,{a}^{2}} }
	+\mathcal{O}\left(\tau^{4}\right)
		\mbox{ for small } \tau.
\end{equation}
Therefore, the central slope or rotation of the simply supported Rayleigh beam of length $L$, with all other parameters set to unity, and with a unit angular impulse acting at \(x=\frac{L}{2},\) is
\begin{multline}\label{slope_fun1}
	u_{,x}\left(x=\frac{L}{2},\tau\right)=\frac{2}{L}\,S(\tau)=\frac{2\,a}{\pi}\,S(\tau)=
	{\frac {1}{2}}-{\frac {{\rm coth} \left({\frac {\pi}{2\, a}}\right)
			}{2}}\,\tau+{\frac {3}{8}\,{\tau}^{2}}+
		{ \frac{\left( \pi-5\, a\sinh \left({\frac {\pi}{a}} \right) \right)
			}{ {\left(\cosh\left(\frac{\pi}{a}\right)-1\right) } } \frac {{\tau}^{3}}{24
				\,{a}} }
		\\+\mathcal{O}\left(\tau^{4}\right) 
		\mbox{ for small } \tau.
\end{multline}
For long beams, \(a=\frac{\pi}{L}\rightarrow 0 \), and the response any point \(x\) is asymptotically given by
\begin{equation}
	u_{,x}(x,\tau)=\frac{1}{2}-\frac{\tau}{2}+\frac{3}{8}\tau^2-\frac{5}{24}\tau^3+{\cal O}\left(\tau^4\right).
\end{equation}
Now, if a less violent moment $M(\tau)$ acts there for some short time interval, then the angular rotation at that location will be given by the convolution integral
\begin{equation}\label{conv1}
\frac{2}{L} \int_0^{\tau} M(\tau - \xi) \,S(\xi) \, {\rm d} \xi.
\end{equation}
In particular, if $M(\tau) = M_0$, a constant, then we have
\begin{equation}\label{conv2}
\frac{2M_0}{L} \int_0^{\tau} S(\xi) \, {\rm d} \xi = M_0 \left ( \frac{\tau}{2} - \frac{\coth \left ( \frac{\pi}{2\,a} \right )}{4} \tau^2 + \cdots \right ).
\end{equation}
For a long beam, $ \coth\left(\frac{\pi}{2\,a}\right)\rightarrow 1 $, and we have
\begin{equation}
	M_0 \left ( \frac{\tau}{2} - \frac{\coth \left ( \frac{\pi}{2\,a} \right )}{4} \tau^2 + \cdots \right )\rightarrow M_0 \left ( \frac{\tau}{2} - \frac{1}{4} \tau^2 + \cdots \right ).
\end{equation}
We note that the response to a suddenly applied constant moment causes a sudden change in angular velocity but not in rotation, and so it involves bounded strains in the beam. 
\section{Numerical verification}

\begin{figure}[h]
	\centering
	\includegraphics[width=\linewidth]{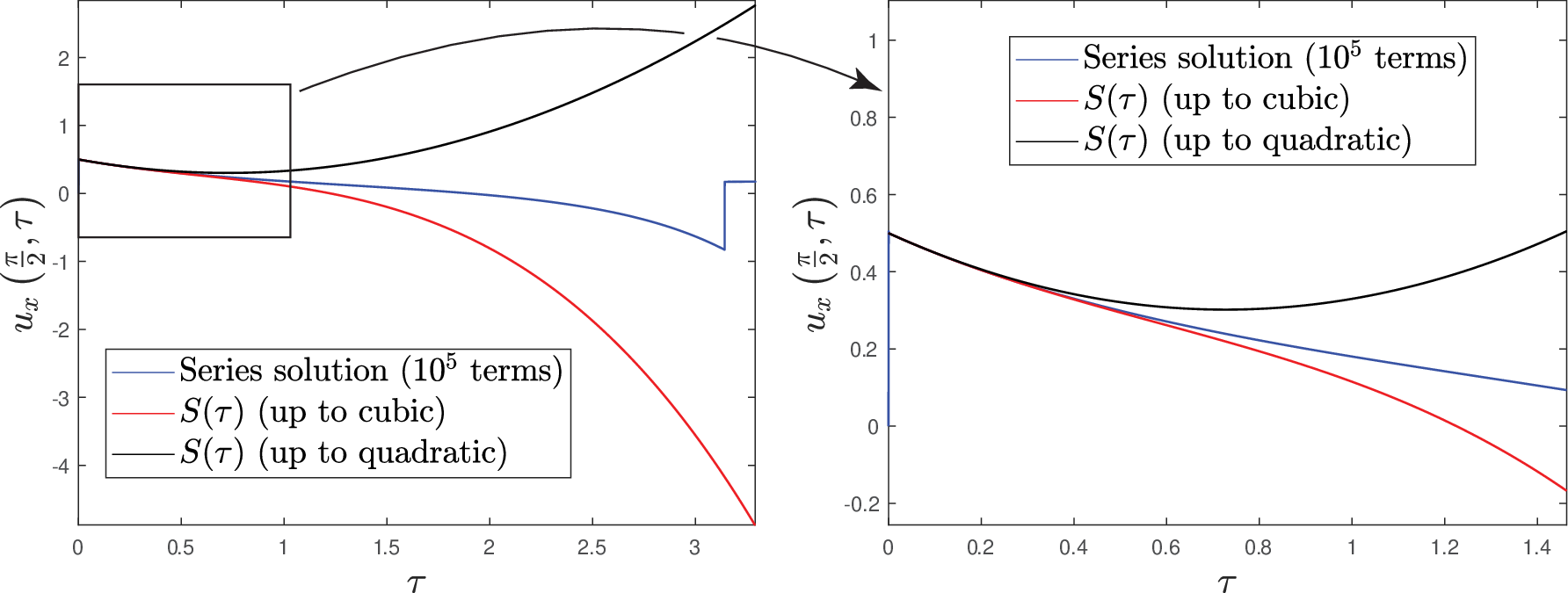}
	\caption{The series solution (\(10^5\) terms) of \(u_x\left(\frac{L}{2},\tau\right)\) and its asymptotic approximation for small \(\tau\) when \(L=\pi\). Both plots
show both quadratic and cubic approximations. The plot on the right shows a smaller time interval to enable clearer comparison for small $\tau$.}
	\label{comparison}
\end{figure}

The accuracy of the asymptotic approximation in Eq.\ \ref{slope_fun1} is displayed in Fig.\ \ref{comparison}. It is clear that the match is good for small $\tau$.

\begin{figure}[h]
	\centering
	\includegraphics[width=0.5\linewidth]{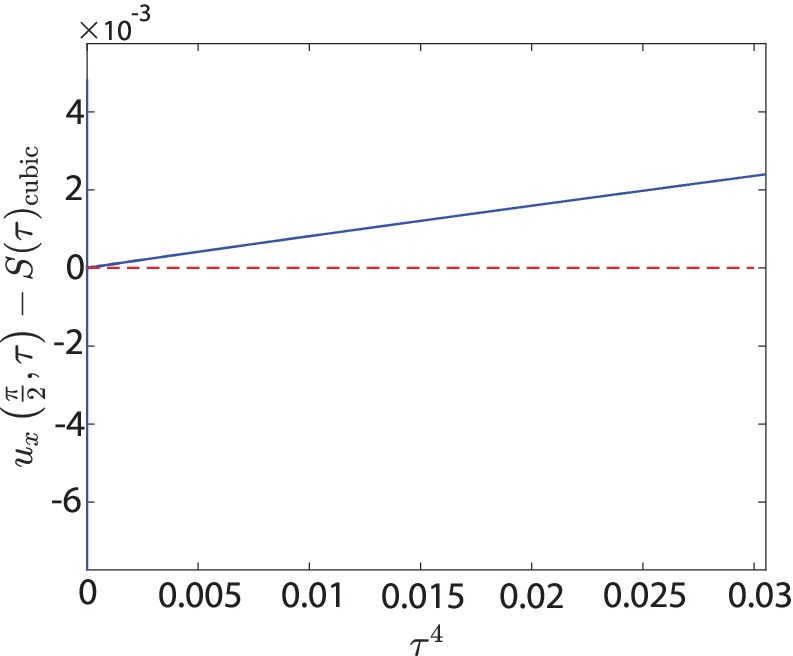}
	\caption{Numerical verification of \(S(\tau)\) up to cubic term.}
	\label{verification}
\end{figure}

To verify numerically that all terms computed in Eq.\ \ref{slope_fun1} are correct, we subtract the asymptotic expression from the series solution
(with $10^5$ terms). The remaining error term should be proportional to $\tau^4$. A plot of that error against $\tau^4$ for small $\tau^4$
should be close to a straight line passing through zero.
Such is indeed the case, as seen in Fig.\ \ref{verification}. The spike seen for super-small $\tau$ is actually not from the asymptotic approximation but from fast
Gibbs oscillations in the truncated series solution. We conclude that, from the numerical evidence, the series computed indeed is correct up to ${\cal O}(\tau^3)$.

\section{Results and discussion}
The prior literature contains several papers on the short time response of some slender structures to {\em linear} impulses, or impulsive forces, concentrated at points in
space. However, the rotational response of slender structures to angular impulses has not been studied. The simplest slender structure, namely the Euler-Bernoulli beam, predicts strongly unphysical responses to angular impulses. The infinite beam has an instantaneously unbounded rotational response, while a finite-length beam shows strong arrivals of high frequency reflections from the boundary.

In contrast to the Euler-Bernoulli beam (which does not incorporate cross-sectional rotary inertia), the Rayleigh beam incorporates rotary inertia and yet retains simplicity because it uses the same kinematics. This gives more physically reasonable results under impulsive moments.
Interestingly the Rayleigh beam and an Eringen stress gradient theory based beam are governed by the same equation. A study of such structures under impulsive moments can have some practical value in that it may shed light on peculiarities observed in some experimental situations like long beams on which a motor is suddenly braked to a halt. However, in this paper our interest has been primarily academic. Using dimensional analysis followed by asymptotic approximations, we have computed the short-time response of such a beam to an angular impulse. The response is found to contain an instantaneous finite jump (unlike an unbounded jump for the Euler-Bernoulli beam), followed by regular behavior well described by a polynomial in nondimensional time for a finite duration, until simultaneous arrival of multi-frequency reflections from boundaries causes further discontinuities in the response. 
We note that, although the rotational response to the angular impulse is bounded as \(t\rightarrow 0\), the resulting strain in the beam is unbounded. As a result, for realistic modeling of beams with such an angular impulse, a nonlinear material model will be needed. However, if the actual moment applied is in fact a smooth function of time, and the impulse response computed above is used merely as a Green's function, then finite responses with finite strains will occur. Recall, e.g., Eq.\ \ref{conv2}. With these thoughts, we acknowledge that nonlinearity may play a role in some situations with some loadings for some problems. But our basic elastic solution has its own fundamental theoretical validity, and is the main contribution of this paper.

We have investigated this problem using dimensional analysis, modal expansion, series solutions, finite element simulations, and asymptotic approximations for the short-time polynomial behavior. Agreement between solutions obtained using different means has been excellent. This work provides new theoretical results for a problem that has not been examined before. These results shed some light on some peculiarities of the response as explained above. Moreover, we hope that that methods used here will be of interest to a general readership interested in both asymptotic analyses as well as structural vibrations.

\section{Acknowledgment}

We thank two anonymous reviewers whose comments led to significant improvements.

\appendix

\section{Finite element analysis}\label{FEA}
For the finite element analysis of the Rayleigh beam, the effect of rotary inertia is to be incorporated in the mass matrix. However, the stiffness matrix remains same as that of Euler-Bernoulli beams. For the \(i^{\rm th}\) element of size \(\ell_{(i)}\), the element mass and stiffness matrices are given by

\begin{equation}\label{FE_element_mass_matrix}
	{\bf M}_{(i)}=\frac{\rho A}{420}\, \left[ \begin {array}{cccc} 156\,{\ell_{(i)}}&22\,\ell_{(i)}^{2}&54\,{
		\ell_{(i)}}&-13\,\ell_{(i)}^{2}\\ \noalign{\medskip}22\,\ell_{(i)}^{2}&4\,
	\ell_{(i)}^{3}&13\,\ell_{(i)}^{2}&-3\,\ell_{(i)}^{3}
	\\ \noalign{\medskip}54\,{\ell_{(i)}}&13\,\ell_{(i)}^{2}&156\,{\ell_{(i)}}&-22
	\,\ell_{(i)}^{2}\\ \noalign{\medskip}-13\,\ell_{(i)}^{2}&-3\,\ell_{(i)}^{3}&-22\,\ell_{(i)}^{2}&4\,\ell_{(i)}^{3}\end {array} \right] 
	+\frac{\rho I}{30}\,
	\left[ \begin {array}{cccc} 36&3\,{\ell_{(i)}}&-36&3\,{\ell_{(i)}}
	\\ \noalign{\medskip}3\,{\ell_{(i)}}&4\,\ell_{(i)}^{2}&-3\,{\ell_{(i)}}&-
	\ell_{(i)}^{2}\\ \noalign{\medskip}-36&-3\,{\ell_{(i)}}&36&-3\,{\ell_{(i)}}
	\\ \noalign{\medskip}3\,{\ell_{(i)}}&-\ell_{(i)}^{2}&-3\,{\ell_{(i)}}&4\,
	\ell_{(i)}^{2}\end {array} \right]
\end{equation}

\begin{equation}\label{FE_element_stiffness_matrix}
	{\bf K}_{(i)}=\frac{EI}{\ell_{(i)}^3}\,
	\left[ \begin {array}{cccc} 12&6\,{\ell_{(i)}}&-12&6\,{\ell_{(i)}}
	\\ \noalign{\medskip}6\,{\ell_{(i)}}&4\,\ell_{(i)}^{2}&-6\,{\ell_{(i)}}&2\,
	\ell_{(i)}^{2}\\ \noalign{\medskip}-12&-6\,{\ell_{(i)}}&12&-6\,{\ell_{(i)}}
	\\ \noalign{\medskip}6\,{\ell_{(i)}}&2\,\ell_{(i)}^{2}&-6\,{\ell_{(i)}}&4\,
	\ell_{(i)}^{2}\end {array} \right]
\end{equation}

The assembly of the global mass (\({\bf M}\)) and stiffness (\({\bf K}\)) matrices is routine  \cite{cook2000fea}. We consider \(n_{\rm e}\) equal-length elements
for simplicity.
The assembled non-homogeneous equations are of the form
\begin{equation}\label{global_FE_equations}
	{\bf M}\,\ddot{\boldsymbol{y}}+{\bf K}\,\boldsymbol{y}=\boldsymbol{f},
\end{equation}
where \(\boldsymbol{y}\) is the global nodal coordinate vector. In our problem the forcing vector \(\boldsymbol{f}\) has all elements equal to zero except \(\delta(\tau)\) (Dirac delta function) at the nodal entry corresponding to the rotational degree of freedom at \(x=\frac{L}{2}\). Alternatively, we can formulate an equivalent homogeneous equation of motion   
\begin{equation}\label{global_FE_equations1}
	{\bf M}\,\ddot{\boldsymbol{y}}+{\bf K}\,\boldsymbol{y}={\bf 0}
\end{equation}
with initial conditions
\begin{equation}\label{initial_conditions}
	\boldsymbol{y}(0)=\boldsymbol{0},\qquad \mbox{and}\qquad \dot{\boldsymbol{y}}(0)={\bf M}^{-1}\,\hat{\boldsymbol{f}}
\end{equation}
where the vector \(\hat{\boldsymbol{f}}\) has all elements zero except  `1' at the nodal entry corresponding to the rotational degree of freedom at \(x=\frac{L}{2}\) .
For time integration of Eq.\ \ref{global_FE_equations1}, we have used the implicit time marching algorithm of Pich\'e \cite{piche1995stable}. We mention that that algorithm has given
very good results in some other recent work with slender structures \cite{goswami2023semi}.

\section{Proof of Lemma 1}\label{app1}
Let us consider a real valued, several-times differentiable function \(g\) with the properties \(g(x)\rightarrow 0\) as \(x\rightarrow \infty\). Moreover, successive derivatives of \(g\) go to zero faster and faster as \(x\rightarrow \infty\). Then
\begin{align}
	\int_{N}^{\infty}g(y)\;{\rm d}y&=\sum_{k=N}^{\infty}\int_{k}^{k+1}g(y)\;\dd y \nonumber \\
	&\sim\sum_{k=N}^{\infty}\int_{k}^{k+1}\left(g(k)+(y-k)g'(k)+\frac{1}{2}(y-k)^2g''(k)+\frac{1}{6}(y-k)^3g'''(k)+\frac{1}{24}g''''(k)(y-k)^4\right)\;\dd y \nonumber\\
	&=\sum_{k=N}^{\infty} g(k)+\frac{1}{2}\sum_{k=N}^{\infty}g'(k)+\frac{1}{6}\sum_{k=N}^{\infty}g''(k)+\frac{1}{24}\sum_{k=N}^{\infty}g'''(k)+\frac{1}{120}\sum_{k=N}^\infty g''''(k). \nonumber
\end{align}
Rearranging,
\begin{equation} \label{gsum}
\sum_{k=N}^{\infty} g(k) = \int_{N}^{\infty}g(y)\;{\rm d}y-\frac{1}{2}\sum_{k=N}^{\infty}g'(k)-\frac{1}{6}\sum_{k=N}^{\infty}g''(k)-\frac{1}{24}\sum_{k=N}^{\infty}g'''(k)-\frac{1}{120}\sum_{k=N}^\infty g''''(k) + \cdots.
\end{equation}
The same equation can be used recursively on the sums on the right hand side; the advantage in such cases is that the antiderivatives are obvious.
More terms can be retained easily if we wish. In this way, we obtain Eq.\ \ref{Lem1}.

\section{Infinite series and their sum}\label{app2}
Some infinite series are listed below. They can be obtained using some tricks with Fourier series; and they can also be obtained from symbolic algebra packages like Maple. Proofs are omitted.
\begin{align}
	&\sum_{k=1}^{\infty}\frac{1}{1+4\,k^2\,a^2}={\frac {1}{4\,a} \left( -2\,a+\pi\,{\rm coth} \left({\frac {\pi}{2\,a}
		}\right) \right) }\\
	&\sum_{k=1}^{\infty}\frac{1}{(1+4\,k^2\,a^2)^2}={\frac {1}{16\,{a}^{2}} \left( {\pi}^{2} \left( {\rm coth} \left({
			\frac {\pi}{2\,a}}\right) \right) ^{2}-8\,{a}^{2}+2\,\pi\,{\rm coth} 
		\left({\frac {\pi}{2\,a}}\right)a-{\pi}^{2} \right) }\\
	&\sum_{k=1}^{\infty}\frac{1}{(1+4\,k^2\,a^2)^3}=\frac {1}{64\,{a}^{3}} \left(  \left( {\rm coth} \left({\frac {\pi}{2
				\,a}}\right) \right) ^{3}{\pi}^{3}
			+3\, \left( {\rm coth} \left({
			\frac {\pi}{2\,a}}\right) \right) ^{2}{\pi}^{2}a-32\,{a}^{3}-{\rm coth} 
		\left({\frac {\pi}{2\,a}}\right){\pi}^{3}+6\,\pi\,{\rm coth} \left({\frac {\pi}{2\,a}}\right){a}^{2}-3\,{\pi}^{2}a \right) 
\end{align}
\section{Derivation of the governing equation}

The kinetic energy of the beam is \cite{hagedorn2007wave} 
\begin{equation}
	{\cal T}=\half\int_{0}^L \rho\, A\, u_{,t}^2 \,\dd x+\half \int_{0}^L \rho\, I \,u_{,xt}^2\,\dd x,
\end{equation}
the potential energy due to bending is
\begin{equation}
	{\cal V}=\half\int_{0}^L EI \,u_{,xx}^2\, \dd x,
\end{equation}
and the work done by the  nonconservative moment
\begin{equation}
	{\cal W}_{\rm nc}=\int_{0}^L M(x,t)\,u_{,x}\, \dd x
\end{equation}
where \(M(x,t)\) is the applied moment. For a beam of length \(L\) with an impulsive moment \(M_0\) at the midpoint, 
\begin{equation}\label{The_moment}
	M(x,t)=M_0\,\delta\left(x-\frac{L}{2}\right)\delta(t).
\end{equation}

For an arbitrary variation, say \(\delta u\), of \(u\), using the extended Hamilton's principle, we have
\begin{equation}\label{eh_ST}
	\int_{t_1}^{t_2}\left(\delta{\cal T}-\delta{\cal V}+\delta{\cal W}_{\rm nc}\right) \, \dd t=0.
\end{equation}
The variation \(\delta u\) as well as \(\delta u_{,x}\) both identically vanish at \(t=t_1\) and \(t=t_2\).
Upon integration by parts, the first term of Eq.\ \ref{eh_ST}
\begin{align}\label{eh_T1}
	\int_{t_1}^{t_2}\delta{\cal T}\,\dd t=&\int_{0}^L\int_{t_1}^{t_2}\left(\rho \, A\, u_{,t}\,\delta u_{,t}+\rho\, I\,u_{,xt}\,\delta u_{,xt} \right)\dd t\,\dd x\nonumber\\
	=&\cancelto{0}{\int_{0}^L \rho \, A \,u_{,t}\,\delta u\bigg\rvert_{t_1}^{t_2}\,\dd x}-\int_{0}^L\int_{t_1}^{t_2} \rho\,A\, u_{,tt}\,\delta u \,\dd x\dd t+\cancelto{0}{\int_{0}^L\rho\,I\,u_{,xt}\,\delta u_{,x}\bigg\rvert_{t_1}^{t_2}\dd x}-\int_{0}^L\int_{t_1}^{t_2}
	\rho\,I\,u_{,xtt}\,\delta u_{,x}\,\dd x\dd t\nonumber\\
	=&-\int_{0}^L\int_{t_1}^{t_2} \rho\,A\, u_{,tt}\,\delta u \,\dd x\dd t-\underset{\small{\mbox{Boundary term}}}{\int_{t_1}^{t_2}\rho\, I\,u_{,xtt}\,\delta u\bigg\rvert_0^L \dd t}+\int_{0}^L\int_{t_1}^{t_2}\rho\,I\,u_{,xxtt}\,\delta u\,\dd x\,\dd t.
\end{align}
Similarly, the second term of Eq.\ \ref{eh_ST} takes the form
\begin{align}\label{eh_T2}
	\int_{t_1}^{t_2}-{\cal V}\,\dd t=&-\int_{0}^L\int_{t_1}^{t_2}EI\,u_{,xx}\delta u_{,xx}\,\dd x\,\dd t\nonumber\\
	=&\underset{\small{\mbox{Boundary term}}}{-\int_{t_1}^{t_2}EI\, u_{,xx}\,\delta u_{,x}\bigg\rvert_0^L\dd t}+\underset{\small{\mbox{Boundary term}}}{\int_{t_1}^{t_2}EI\,u_{,xxx}\,\delta u\bigg\rvert_0^L\dd t}-\int_{0}^L\int_{t_1}^{t_2}EI\,u_{,xxxx}\,\delta u\,\dd x\,\dd t,
\end{align}

and the third term of Eq.\ \ref{eh_ST} results in
\begin{equation}\label{eh_T3}
	\int_{t_1}^{t_2}\delta {\cal W}_{\rm nc}\,\dd t=\int_{0}^L\int_{t_1}^{t_2}M(x,t)\delta u_{,x}\,\dd x\,\dd t=\underset{\small{\mbox{Boundary term}}}{\int_{t_1}^{t_2}M(x,t)\,\delta u \bigg\rvert_0^L \dd t}-\int_{0}^L\int_{t_1}^{t_2}\pdf{M(x,t)}{x}\delta u \,\dd x\,\dd t.
\end{equation}
Combining Eqs.\ \ref{eh_T1}, \ref{eh_T2}, and \ref{eh_T3}, and invoking the fundamental lemma of calculus of variations \cite{hagedorn2007wave}, we obtain
\begin{equation}
	\rho \, A\,u_{,tt}+EI\,u_{,xxxx}-\rho\,I\,u_{,xxtt}=-\pdf{M(x,t)}{x}.
\end{equation}
Note that the above equation is accompanied by boundary conditions. We have considered simply supported boundary conditions in our study. Recalling Eq.\ \ref{The_moment}, we obtain Eq.\ \ref{gov_eq}, i.e.,
\begin{equation}
	\rho \, A\,u_{,tt}+EI\,u_{,xxxx}-\rho\,I\,u_{,xxtt}=-M_0\, \delta_{,x}\left(x-\frac{L}{2}\right)\,\delta(t).
\end{equation}


%


\begin{thebibliography}{25}

\bibitem{timoshenko2003strength}Timoshenko, S. P., 2003, \emph{History of Strength of Materials}, Dover, USA.   

\bibitem{graff2012wave}Graff, K. F., 1975, \emph{Wave Motion in Elastic Solids}, Oxford University Press, Oxford, UK.  

\bibitem{rayleigh1945tos}Rayleigh, J. W. S., 1945, \emph{Theory of Sound}, Dover, USA.  

\bibitem{timoshenko1921model}Timoshenko, S. P., 1921, On the correction for shear of the differential equation for transverse vibrations of prismatic bars, \emph{Philosophical Magazine}, 41(245): 744-746.  

\bibitem{timoshenko1922model}Timoshenko, S. P., 1922, On the transverse vibrations of bars of uniform cross-section, {\em The London, Edinburgh, and Dublin Philosophical Magazine and Journal of Science}, 43(253): 125-131.  

\bibitem{meirovitch1997vibrations}Meirovitch, L., 1997, \emph{Principles and Techniques of Vibrations}, Prentice-Hall, USA.   

\bibitem{chatterjee2004short}Chatterjee, A., 2004, The short-time impulse response of Euler-Bernoulli beams, \emph{Journal of Applied Mechanics, ASME}, 71(2): 208-218.

\bibitem{zener1941intrinsic}Zener, C., 1941, The intrinsic inelasticity of large plates, \emph{Physical Review}, 59(8): 669. 

\bibitem{schwieger1965simple}Schwieger, H., 1965, A simple calculation of the transverse impact on beams and its experimental verification, {\em Experimental Mechanics}, 5(11): 378-384.

\bibitem{schwieger1970central}Schwieger, H., 1970, Central deflection of a transversely struck beam, \emph{Experimental Mechanics}, 10(4): 166-169. 

\bibitem{meijaard2007lateral}Meijaard, J., 2007, Lateral impacts on flexible beams in multibody dynamics simulations, \emph{In IUTAM Symposium on Multiscale Problems in Multibody System Contacts, Springer, Dordrecht}, 173-182.

\bibitem{bhattacharjee2018transverse}Bhattacharjee, A., and Chatterjee, A., 2018, Transverse impact of a Hertzian body with an infinitely long Euler-Bernoulli beam, {\em Journal of Sound and Vibration}, 429: 147-161.

\bibitem{claeyssen2002impulse}Claeyssen, J. R., Chiwiacowsky, L. D., and Suazo, G. C., 2002, The impulse response in the symbolic computing of modes for beams and plates, \emph{Applied Numerical Mathematics}, 40(1-2): 119-135.

\bibitem{roy1995transient}Roy, P. K., and Ganesan, N., 1995, Transient response of a cantilever beam subjected to an impulse load, \emph{Journal of Sound and Vibration}, 183(5): 873-880.



\bibitem{barkanov2000transient}Barkanov, E., Rikards, R., Holste, C., and Täger, O., 2000, Transient response of sandwich viscoelastic beams, plates, and shells under impulse loading, \emph{Mechanics of Composite Materials}, 36(3): 215-222.

\bibitem{jayaprakash2013fatigue}Jayaprakash, K., Desai, Y. M., and Naik, N. K., 2013, Fatigue behavior of \([0_n/90_n]_s\) composite cantilever beam under tip impulse loading, \emph{Composite Structures}, 99: 255-263.

\bibitem{wagg1999experimental}Wagg, D. J., Karpodinis, G., and Bishop, S. R., 1999, An experimental study of the impulse response of a vibro-impacting cantilever beam, \emph{Journal of Sound and Vibration}, 228(2): 243-264. 

\bibitem{bhattacharjee2020restitution}Bhattacharjee, A., and Chatterjee, A., 2020, Restitution modeling in vibration-dominated impacts using energy minimization under outward constraints, \emph{International Journal of Mechanical Sciences}, 166: 105215.

\bibitem{kenny2000dynamic}Kenny, S., Pegg, N., and Taheri, F., 2000, Dynamic elastic buckling of a slender beam with geometric imperfections subject to an axial impulse, \emph{Finite Elements in Analysis and Design}, 35(3): 227-246.








\bibitem{langhaar1951}Langhaar, H. L., 1951, \emph{Dimensional Analysis and Theory of Models}, John Wiley \& Sons (reprinted in 1987 by the Robert E. Krieger Publishing Company, Malabar, FL).

\bibitem{hagedorn2007wave}Hagedorn, P., and Dasgupta A., 2007, \emph{Vibration and Waves in Continuous Mechanical Systems}, John Wiley, West Sussex, England.

\bibitem{gopalakrishnan2017wave}Gopalakrishnan, S., 2017, \emph{Wave Propagation in Materials and Structures}, CRC Press, USA.

\bibitem{eringen2002nonlocal}Eringen, A. C., 2002, \emph{Nonlocal Continuum Field Theories}, Springer, USA.

\bibitem{bender1999advanced}Bender, C. M., and Orszag, S., 1999, {\em Advanced Mathematical Methods for Scientists and Engineers: Asymptotic Methods and Perturbation Theory}, Springer Science and Business Media.

\bibitem{cook2000fea}Cook, R. D., Malkus, D. S., Plesha, M. E., 2000, \emph{Concepts and Applications of Finite Element Analysis}, John Wiley, USA.

\bibitem{piche1995stable}Pich\'e, R., 1995, An L-stable Rosenbrock method for step-by-step time integration in structural dynamics, \emph{Computer Methods in Applied Mechanics and Engineering}, 126(3-4): 343-354.

\bibitem{goswami2023semi}Goswami, B., and  Chatterjee, A., 2023, Semi-implicit integration and data-driven model order reduction in structural dynamics with hysteresis, {\em Journal of Computational and Nonlinear Dynamics, ASME}, 18(5): 051002.
	
\end{thebibliography}
\end{document}